\documentclass[
 reprint,
 amsmath,
 amssymb,
aps,
 prb,
showpacs
]{revtex4-1}
\usepackage{amsmath}
\usepackage{amssymb} 
\usepackage{graphicx}
\usepackage{dcolumn}
\usepackage{bm}
\usepackage{hyperref}
\usepackage[abbreviations,detect-mode,separate-uncertainty]{siunitx}

\newcommand{\ket}[1]{\left \vert #1 \right \rangle}

\begin{document}

\preprint{APS/123-QED}

\title{Polarization-assisted Vector Magnetometry in Zero Bias Field with an Ensemble of Nitrogen-Vacancy Centers in Diamond}

\author{F. M\"unzhuber}
\author{J. Kleinlein}
\author{T. Kiessling} \email[]{tobias.kiessling@physik.uni-wuerzburg.de}
\author{L. W. Molenkamp}
\affiliation{Physikalisches Institut Universit\"at W\"urzburg, Am Hubland, 97074 W\"urzburg, Germany}%

\begin{abstract}

We demonstrate vector magnetometry with an ensemble of nitrogen-vacancy (NV) centers in diamond without the need for an external bias field. The anisotropy of the electric dipole moments of the NV center reduces the ambiguity of the optically detected magnetic resonances upon polarized visible excitation. Further lifting of the remaining ambiguities is achieved via application of an appropriately linearly polarized microwave field, which enables suppression of spin-state transitions of a certain crystallographic NV orientation. This allows for the full vector reconstruction of small ($\leq \SI{0.1}{\milli\tesla}$) magnetic fields without an external bias field having to interfere with the magnetic structure. 

\end{abstract}


\maketitle

\section{Introduction}

The negatively charged nitrogen-vacancy (NV) center in diamond has attracted enormous research interest in the last decade because of its formidable versatility in photonic applications. In addition to its suitability as quantum information processing tool~\cite{Childress13102006}, quantum cryptography element~\cite{PhysRevLett.85.290, PhysRevA.64.061802} and frequency standard~\cite{PhysRevA.87.032118}, the defect is a prospective candidate for the realization of quantum sensors for a wide band of physical parameters, such as temperature~\cite{PRLbudkerD(T), lukind(t)}, acceleration~\cite{gyroscope}, pressure~\cite{PRLpressure}, electric~\cite{Dolde2011}, and in particular magnetic fields~\cite{Balasubramanian2008, Maze2008}. The single atom-like defect embedded in a controllable, but nearly non-interacting solid state environment enables the construction of magnetometers with an unprecedented combination of spatial and magnetic field resolution. Previously proposed schemes build on established scanning probe technology and add the magnetic field sensitivity of an atomic gas sensor, which promises insight into new physics~\cite{scannanores, scannanoresII, apldegen, budker2013optical}.

Magnetometry with NV centers is based on optically detected magnetic resonance (ODMR) spectroscopy of the field-sensitive spin states of the defect~\cite{JoPODMR, 0034-4885-77-5-056503}. The multiplicity of the electronic ground state of the NV center is $S=1$. The three resulting eigenstates along the quantization axis, which is set by the crystallographic orientation of the NV axis, are usually labeled as $\ket{m_S = -1,0,1}$. Due to crystal field splitting the $\ket{\pm 1}$-states are energetically shifted from the $\ket{0}$-state. Optical excitation of the center as well as subsequent optical recombination are spin conserving. As crucial ingredient, the optical intensity of the $\ket{\pm 1}$-states is weaker compared to the $\ket{0}$-state because of an additional recombination channel in the infrared for this configuration~\cite{APLinfrared}. Therefore, the intensity of the optical photoluminescence (PL) signal can serve as a measure for the NV spin state.

For an external magnetic field $B_{ext}$ parallel or antiparallel to the NV axis, the energetic shift of the states is described by the Zeeman term

\begin{align}
\Delta E_{Zeeman} =g \mu_B B_{ext} S_z
\label{eq:zeeman}
\end{align}
where $g$ is the electron g-factor, $\mu_B$ the Bohr magneton and $S_z$ the projection of the spin along the quantization axis. (A general description for arbitrarily oriented fields can be found e.g. in Ref.\cite{Balasubramanian2008}.) Accordingly, only the energetic position of $\ket{\pm 1}$-states changes as a function of the external field. 

Direct transitions from $\ket{0} \rightarrow \ket{\pm 1}$ can be introduced via the application of a microwave (MW) field, which matches the splitting between the states in energy. Because the splitting is sensitive to the magnetic field at the site of the NV center, the local magnetic field can be probed via the MW frequency needed to induce an ODMR Signal.  

Two approaches for NV center based magnetometers can be found in the literature. Using a single NV center allows a spatial resolution given by the Bohr radius of the defect, which is on the nm scale~\cite{MaletinskyP.2012, Rondin2013, Grinolds2013, HaeberleT.2015, Luan2015}. In this configuration only the projection of the magnetic field vector along the NV axis can be measured. Determining the other components of the field vector thus requires more than one sensor.

The second approach employs an ensemble of NV centers~\cite{aplawschvec, APLinfrared, Glenn2015}. The NV centers are oriented along the four crystallographic axes within the diamond lattice. A given magnetic field will obviously have different projections along these distinct axes. With all four possible NV orientations present in the observation volume, all vector components can be measured with the same probe in one run. 

\begin{figure}\includegraphics{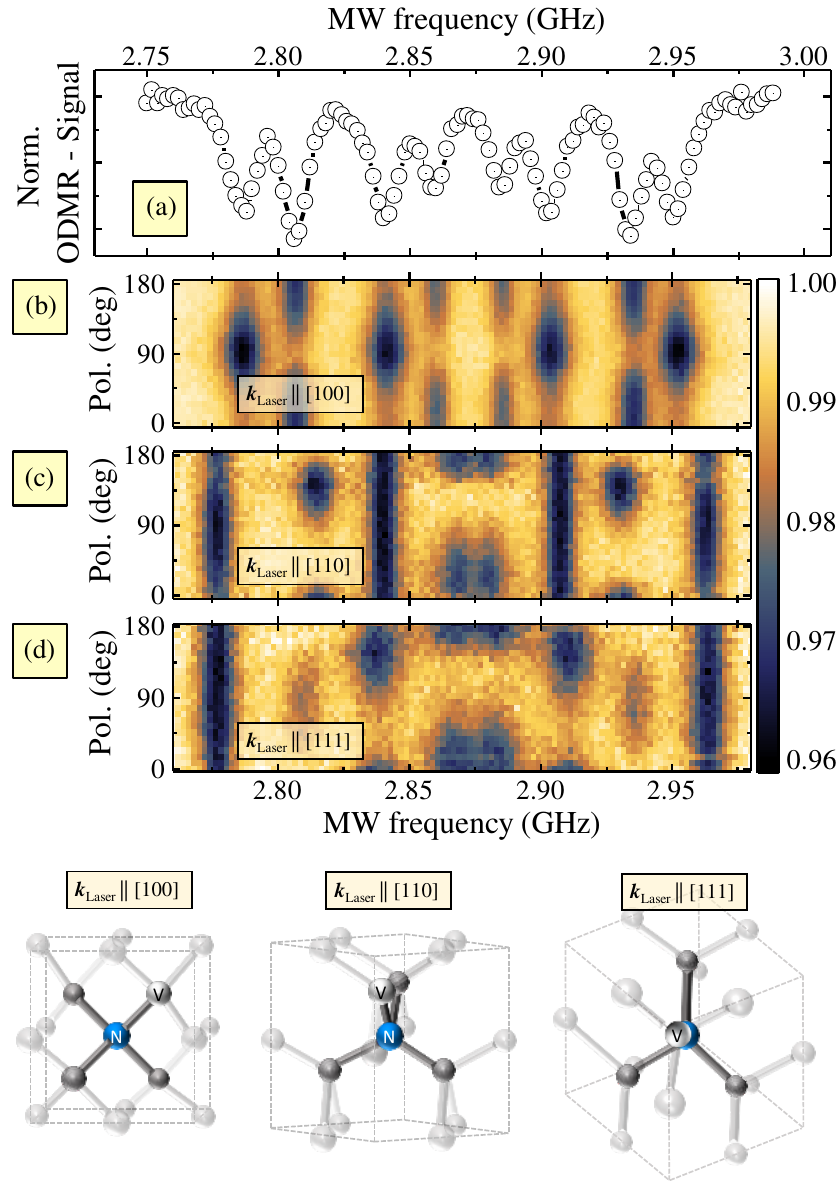}
	\caption{\textbf{(a)} NV ensemble ODMR spectrum in random magnetic field  with random orientation of laser polarization. The field orientation is chosen such that each of the eight possible resonances is clearly resolved. \textbf{(b)} ODMR signal as a function of the MW frequency and the orientation angle of the linear polarization excitation plane. A horizontal cut resembles an ODMR spectrum as in (a). The pairwise appearance of resonances under variation of the polarization angle is characteristic for light incident along the direction of the [100]-crystallographic axis. Magnetic field ($\vert \vec{B} \vert \approx \SI{4}{\milli\tesla}$) has random orientation. \textbf{(c)} Same as (b) for light incident along the [110]-axis. The contribution from two NV orientations can be clearly assigned, the two other remain indistinguishable. Magnetic field is oriented coarsely along [110]. \textbf{(d)} Same as (b)for light incident along the [111]-axis. Each of the four NV orientations gives an unique response to the variation of the laser polarization. Magnetic field is oriented coarsely along [110]. Below: Schematic view of the diamond lattice along the [100]- (left), [110]- (center), and [111]- (right) crystallographic axis for better visualization.}
	\label{fig:01}
\end{figure}

The high information density of the resulting ODMR spectrum is, however, not straightforward to analyze. A bare spectrum as depicted in Fig.~\ref{fig:01} (a)  contains no information on which pair of the resulting eight resonances belongs to which NV orientation. Without further input about the external field, the correct reconstruction of the magnetic field vector is impossible~\cite{revsciinstarray}.

In principle, an additional bias field that has different projections along the four NV axes can be used to identify the resonances~\cite{aplawschvec}. However, it may not always be desirable to expose the investigated sample to a magnetic bias field.

In the following we demonstrate how the ambiguity of the resonances can be lifted by exploiting the NV center selection rules for both the excitation in the visible region and the induced transitions by the MW field. 

\section{Polarization selective visible excitation of NV centers}

The first prerequisite for the appearance of a specific resonance in the ODMR spectrum is optical excitation of the NV centers along the associated crystallographic axis. We use a $\SI{532}{\nano\meter}$ diode laser system to excite the NV centers in a standard confocal geometry. The orientation of the linear polarization plane of the laser beam is tuned by the rotation of a $\lambda / 2$ wave plate in the excitation path. We achieve a spatial resolution of $\SI{700}{\nano\meter}$ by employing an infinity-corrected, long working distance microscope objective and by using a pinhole as spatial filter in the detection path. The sample is placed in the center of a 3D vector magnet on a three dimensional, closed-loop translation stage, which is used to raster scan the sample position. The PL signal from the NV centers is filtered spectrally by a $\SI{690}{\nano\meter} \pm \SI{40}{\nano\meter}$ band pass. 

The excitation efficiency is crucially determined by the coupling strength between the electric field of the laser beam and the dipole moments of electron orbitals of the NV center, which are oriented perpendicular to the NV axis and to each other. For instance, the dipole moments $d1$ and $d2$ of the [111]-oriented NV axis point along [$\overline{2}$11]  and [01$\overline{1}$]~\cite{epstein,PhysRevB.76.165205}.

The probability $\Gamma$ to excite a NV center is proportional to the projection of the electric field vector $\vec{\varepsilon}$ of the incoming laser beam on its dipole moments $\vec{d}$~\cite{gasiorowicz}: 

\begin{align}
\Gamma \propto \left|  \vec{\varepsilon} \cdot \vec{d} \right|^2
\label{eq:fgr}
\end{align}

For laser incidence parallel to the [100] axis, the polarization can be rotated in the (100)-plane via the $\lambda / 2$ wave plate in the excitation path. The field vector $\vec{\varepsilon}$ is then described by

\begin{align}
\vec{\varepsilon} = \begin{bmatrix} 0 \\ E \sin \theta_{pol} \\ E \cos \theta_{pol} \end{bmatrix}
\end{align}
with $\theta_{pol}$ the angle between the polarization and a specific axis in the (100)-plane. Accordingly, the probability $\Gamma$ is a function of the angle of laser polarization:

\begin{align}
\Gamma_{d1} & \propto \left| \begin{bmatrix} 0 \\ E \sin \theta_{pol} \\ E \cos \theta_{pol} \end{bmatrix} \begin{bmatrix} 0 \\ 1/\sqrt{2} \\ -1/\sqrt{2}  \end{bmatrix} \right|^2 \notag \\
& = \frac{E^2}{2} \left( \sin \theta_{pol}-\cos \theta_{pol} \right)^2 = \frac{E^2}{4} \cos \left(\theta_{pol}+\frac{\pi}{4} \right)^2, \notag \\
\notag \\
\Gamma_{d2} & \propto \left| \begin{bmatrix} 0 \\ E \sin \theta_{pol} \\ E \cos \theta_{pol} \end{bmatrix} \begin{bmatrix} -2/\sqrt{6} \\ 1/\sqrt{6} \\ 1/\sqrt{6} \end{bmatrix} \right|^2 \notag \\ 
& = \frac{E^2}{6} \left( \sin \theta_{pol} + \cos \theta_{pol} \right)^2 = \frac{E^2}{12} \sin \left( \theta_{pol}+\frac{\pi}{4} \right)^2, \notag \\
\notag\\
I & \propto \Gamma_{d1} + \Gamma_{d2} \notag \\
&\propto \frac{E^2}{4}  \left(  \cos \left(\theta_{pol}+\frac{\pi}{4} \right)^2 +  \frac{ \sin \left( \theta_{pol}+\frac{\pi}{4} \right)^2}{3} \right).
\label{eq:dipvec}
\end{align}
Consequently, the intensity $I$ of the optical transitions changes upon rotation of the linear polarization of the excitation.

Looking onto a (100)-facet of diamond, one can distinguish (Fig.~\ref{fig:01} below) pairs of the NV axes that have parallel projections into the observational plane. Accordingly, the resonances of these orientations appear and quench pairwise during the rotation of the linear polarization of the exciting laser beam. The number of possible NV orientations to which a resonance can be assigned is therefore reduced to two. Vice versa, a specific orientation of the polarization allows for selective NV center excitation of certain crystallographic orientations.

This principle can be further improved by using samples of diamond that show (110)- or (111)-facets. If the incident laser beam is parallel to the [110]-axis, two of the four orientations are parallel to the observational plane. According to Eq.~\ref{eq:fgr}, the angle of polarization can be chosen such that the signal from those orientations is entirely suppressed. These can, therefore, unambiguously be identified from their response to the variation of the excitation polarization, as shown in Fig.~\ref{fig:01} (c). However, an assignment of the two remaining resonances to the two out-of-plane orientations remains unfeasible.

As demonstrated in Fig.~\ref{fig:01} (d), illumination of the diamond along the [111]-axis allows to overcome this deficit. Three NV orientations are nearly parallel to the observational plane and the angles of laser polarization for which they are preferentially excited are at intervals of $\Delta \theta = \SI{60}{\degree}$. Only the out-of-plane orientation is constantly excited at any angle of polarization, which renders it easy to identify. In principle, all eight resonances can be assigned unequivocally to their corresponding NV orientations, i.e., all components of a magnetic field vector can be reconstructed (except their sign) without the need of a bias magnetic field.

In practice, this is complicated if the magnetic fields have equal or nearly equal projections along the out-of-plane NV axis and one or more additional NV axis. In this specific case and more general for small magnetic fields, the spectral positions of the resonances heavily overlap. This renders a convincing evaluation of the ODMR spectrum very challenging. For this reason, it is highly desirable to further reduce the number NV orientations which contribute to the ODMR spectrum.

\begin{figure}
	\centering
		\includegraphics{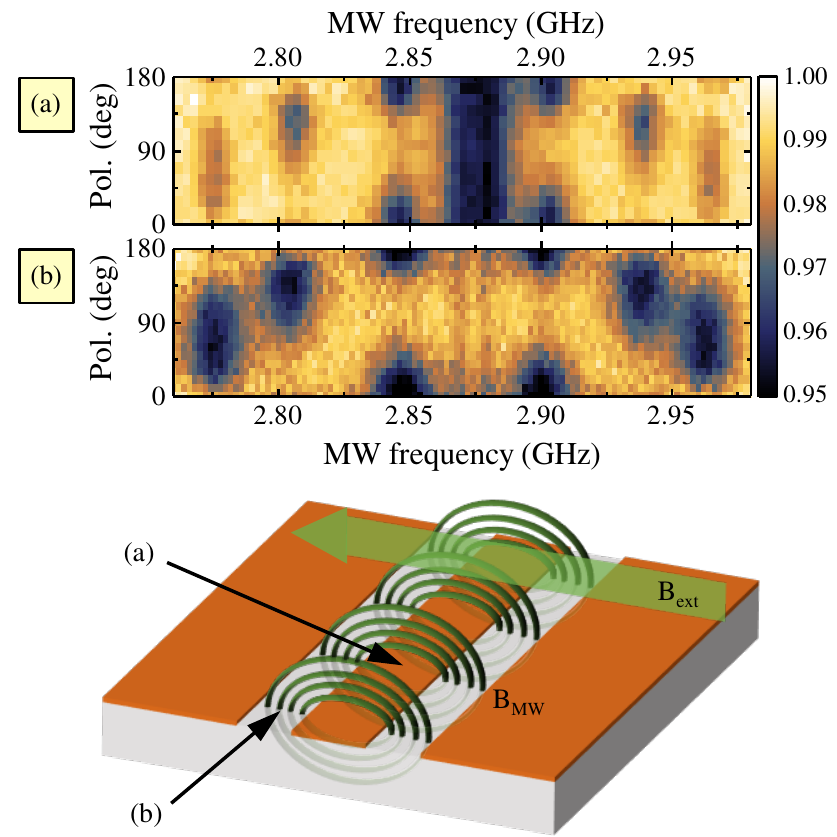}
	\caption{(a), (b) ODMR signal obtained from a [111] oriented diamond with laser illumination along [111]-axis and $\vert \vec{B} \vert \approx \SI{4}{\milli\tesla}$ as a function of the MW frequency and the angle of the linearly polarized excitation laser at different locations above the waveguide. The positions as well as the orientations of the magnetic fields are indicated in the schematic view of the coplanar waveguide below. The material of the waveguide is a standard high frequency circuit material made of ceramic-filled PTFE composites. The width of the signal line is $\SI{500}{\micro\meter}$.}
	\label{fig:02}
\end{figure}





\section{Polarization selective microwave excitation of NV centers}

We can further enhance the directional selectivity by utilizing the selection rules of the microwave-induced NV spin transitions. The Hamiltonian $\mathcal{H}^{MW}$ of the NV center spin subjected to a driving magnetic field $B^{MW}$ of frequency $\omega$ in the presence of a static magnetic field $B_{ext}$ takes the basic structure (in the rotating wave approximation)~\cite{PhysRevA.90.012302}

\begin{equation}
\mathcal{H}_{MW} = \begin{pmatrix} \Delta_- & \epsilon_- & 0\\ \epsilon_-^* & 0 & \epsilon_+ \\ 0 & \epsilon_+^* & \Delta_+ \end{pmatrix}
\label{eq:MWHam}
\end{equation}

where $\Delta_{\pm}=\omega^{\pm}_L{\pm}\omega$ describes the energetic detuning of the MW frequency from the Larmor frequency of the NV spin $\omega_L^{\pm}=D \pm \frac{g \mu_B}{\hbar} B_{ext,axial}$, which itself is set by the crystal field splitting $D$(=2.87~GHz) and the projection $B_{ext}$ of the static magnetic field along the the quantization axis of the NV center (for small fields). The transitions between the $\ket{0}$ and the $\ket{\pm 1}$ states are driven by the $ \epsilon_{\pm} \propto (B^{MW}_x \pm i B^{MW}_y )$ components. In this picture, we immediately recognize that only MW fields which have a $x$- or $y$-component in the NV basis can introduce transitions between the eigenstates and thus cause an ODMR signal.

Guiding the MW via a coplanar waveguide to the diamond, both the magnetic as well as the electric field component of the MW emitted from the waveguide are linearly polarized. The impact of the linear MW polarization is already obvious from the data shown in  Fig.~\ref{fig:01}. Due to the relative orientation of the MW field to the four possible NV axes, the ODMR transitions are noticeably different in amplitude.

To further demonstrate this interplay, we exploit the spatial stray field distribution near the signal lead of the waveguide structure. Depending on whether one probes directly on top of the signal or in the gap between signal and ground lines, the polarization of $B^{MW}$ is either parallel to the waveguide surface or points out of the plane. This dramatically changes the efficiency with which transitions of the out-of-plane NV orientation can be driven.

In Fig.~\ref{fig:02} we present a comparison of ODMR signals acquired at the aforesaid positions. We apply an in-plane external magnetic field in order to evidence the effect on the different transitions. This keeps the contribution of the out-of-plane NV centers at their zero-field value (the physics of the remaining splitting are discussed below). Their spectral position therefore correspond to the situation of a low external field measurement, for which the problem of overlapping resonances occurs.

For polarization parallel to the surface we see in Fig.~\ref{fig:02}~(a) that the out-of-plane orientation dominates the spectrum. The contributions of the remaining in-plane NV orientations are strongly reduced. The weakest signal is observed from the NV orientation that experiences the strongest spectral shift. This is immediately understood from Eq.~\ref{eq:MWHam}. The largest shift corresponds to the largest projection of $B_{ext}$ onto one of the NV axes. As $B_{ext}$ is collinear with the polarization vector of $B^{MW}$ in this configuration, this means that the $B^{MW}_x$ and $B^{MW}_y$ components are minimized. Therefore the MW drives transitions of this NV orientation the least.

The situation changes dramatically for out-of-plane MW fields. The formerly dominating contribution from the  out-of-plane orientation is then virtually eliminated, whereas the in-plane orientations show comparable intensities. 

Combined with the controlled rotation of the exciting laser polarization, we are now in a position to selectively address single NV orientations of an ensemble of NV centers, thereby lifting the inherent ambiguities which otherwise arise when working with an ensemble of NV centers.

\begin{figure}[t]
\includegraphics{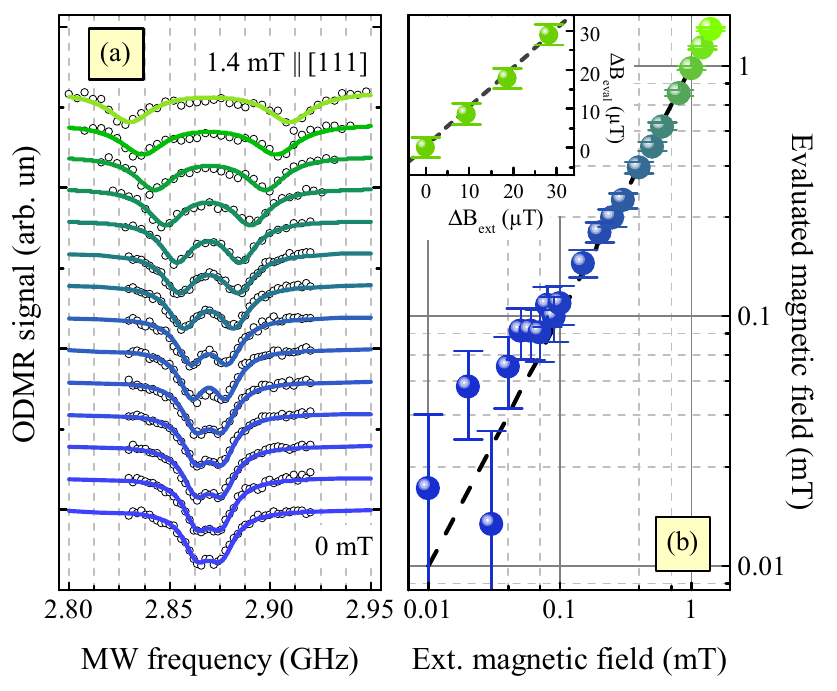}
\caption{(a) Magnetic field response of a single NV orientation, isolated by combination of selective MW and laser excitation. The external field is applied along the [111]-direction. The contribution from the three other NV orientations to the ODMR spectra is negligible. (b) Robustness of the line shape analysis. At small magnetic fields ($B < \SI{50}{\micro\tesla}$) the influence of the transverse crystal field limits the field resolution. The inset shows how the reliability can be further increased by higher frequency sampling, when the transverse crystal field plays a minor role (see text).}
\label{fig:03}
\end{figure}

Before employing this method in a real vector magnetometry experiment, we investigate its sensitivity. To this end, we apply an external magnetic field along a given NV axis and test the accuracy of the line shape analysis.

The response of the selectively excited NV centers is shown in Fig.~\ref{fig:03}~(a). For strong enough external field the spectrum explicitly confirms the excitation of only one orientation, because we can observe merely two transitions instead of four, six, or eight. For very small fields we recognize a further splitting, which is not related to the applied magnetic field and persists even at zero-field value. This splitting is well understood and arises from the transverse component of the crystal field to the NV centers~\cite{PhysRevB.53.13441}.

The experimental spectra in Fig.~\ref{fig:03} (a) can be perfectly reproduced by a model function consisting of two Lorentzians with negative amplitudes. Because of the transverse crystal field splitting the spectral shift of their minima does not follow the external field linearly down to very small magnetic fields~\cite{abragam}. The spectral position of the transitions can be described by

\begin{align}
\Delta \nu = \sqrt{E^2_{trans} + \left( g \mu_B B_{ext} / h \right)^2} 
\label{eq:shift}
\end{align} 
where $E^2_{trans}$ corresponds to the strength of the transverse crystal field.

\begin{figure*}[t]
\includegraphics{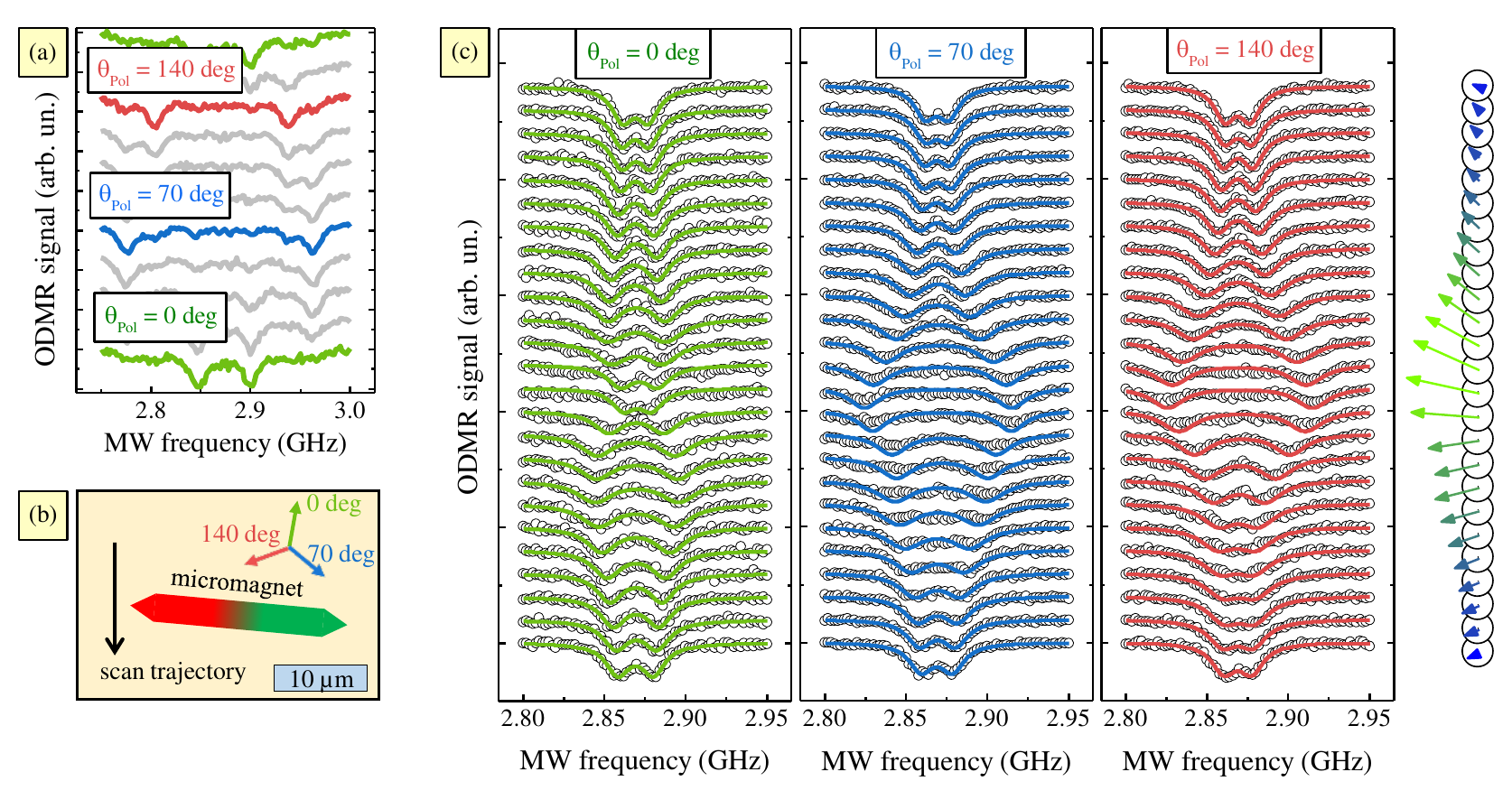}
\caption{(a) ODMR spectra as a function of laser polarization with out-of-plane MW field for suppressing contributions from out-of-plane NV centers. The spectra for each angle of laser polarization are offset for better visualization. The angles with the highest selectivity are $\SI{0}{\degree}$ (green), $\SI{70}{\degree}$ (blue), and $\SI{140}{\degree}$ (red). (b) Schematic view of the diamond surface and the structure on top of it. The scan trajectory, a scale bar, and the relative orientation of the in-plane NV axis to the structure are indicated. The color scheme links the NV orientation to the belonging angle of excitation polarization from (a). (c) ODMR spectra as a function of spatial position and angle of laser polarization. The step size between each measurement point is $\Delta x = \SI{500}{\nano\meter}$. We recognize the individual evolution of the spectra at each angle, confirming the complete selectivity of the excitation. On the right side a reconstruction of the magnetic field vector is shown, reproducing the supposed decay and rotation of the stray field. The length as well as the color scheme corresponds to the field strength. The direction of the arrows corresponds to the direction of the field vector if translated to the trajectory in (b). The size of the circle corresponds to the detection area.}
\label{fig:04}
\end{figure*}

Taking the above into account, our analysis results in a very good agreement between the actual applied field and the field values obtained from the line shape analysis. As a rough estimate, we achieve a field resolution of better than $B = \SI{50}{\micro\tesla}$ at a spectral resolution of $\SI{1}{\mega\hertz}$ and an integration time of $\SI{200}{\milli\second}$. 

From Fig.~\ref{fig:03} (b) one observes a decreasing accuracy and a trend to overestimate the actual field at small fields. This is understood from the structure of Eq.~\ref{eq:shift}.\footnote{The earth field may act as an additional source of error, but this does not alter the following statements.} For small fields, the $E$-part dominates the spectral position and the contribution due to $B_{ext}$ is vanishing. For instance, an external field of $B = \SI{10}{\micro\tesla}$ is expected to result in a shift of $\Delta \nu  = \SI{280}{\kilo\hertz}$ for an undisturbed NV center according to Eq.~\ref{eq:zeeman}. In contrast, having an additional transverse field splitting of $E  = \SI{5}{\mega\hertz}$ as in our diamond (comp. to Fig.~\ref{fig:03} (a)), Eq.~\ref{eq:shift} yields a shift of only $\Delta \nu  = \SI{8}{\kilo\hertz}$.

This effect limits the achievable field resolution at small fields in our measurements. To circumvent this issue, homogenization and relaxation of the local lattice environment are required, which have been demonstrated previously to be feasible~\cite{implantovergrow}. The inset in Fig.~\ref{fig:03} (b) shows that smaller changes in the external field can be resolved, as soon as the resonance shift occurs approximately linearly with the change of the magnetic field. The applied fields range from $\SI{1.20}{\milli\tesla}$ to $\SI{1.23}{\milli\tesla}$ and the sampling rate is set to $\SI{100}{\hertz}$. This enables a field resolution down to a few $\SI{}{\micro\tesla}$.

\section{Spatially resolved vector magnetometry}

As a demonstration of the capabilities of our technique we investigate the stray field of a ferromagnetic stripe structured on the surface of a processed sample of [111] diamond.\footnote{The structure is a $\SI{50}{\nano\meter}$ thick rectangular with equilateral triangles on its short edges made of cobalt. It is deposited directly on the implanted (111)-surface of the diamond. Implantation parameters were chosen to obtain an average implantation depth of $\SI{10}{\nano\meter}$ and a density of ca. $\SI{500}{NV\per\micro\square\meter}$} The geometry of the structure is chosen such that change in amplitude and direction of the static B-field vector occur on a length scale comparable to our resolution limit.

\begin{figure*}[ht]
\includegraphics{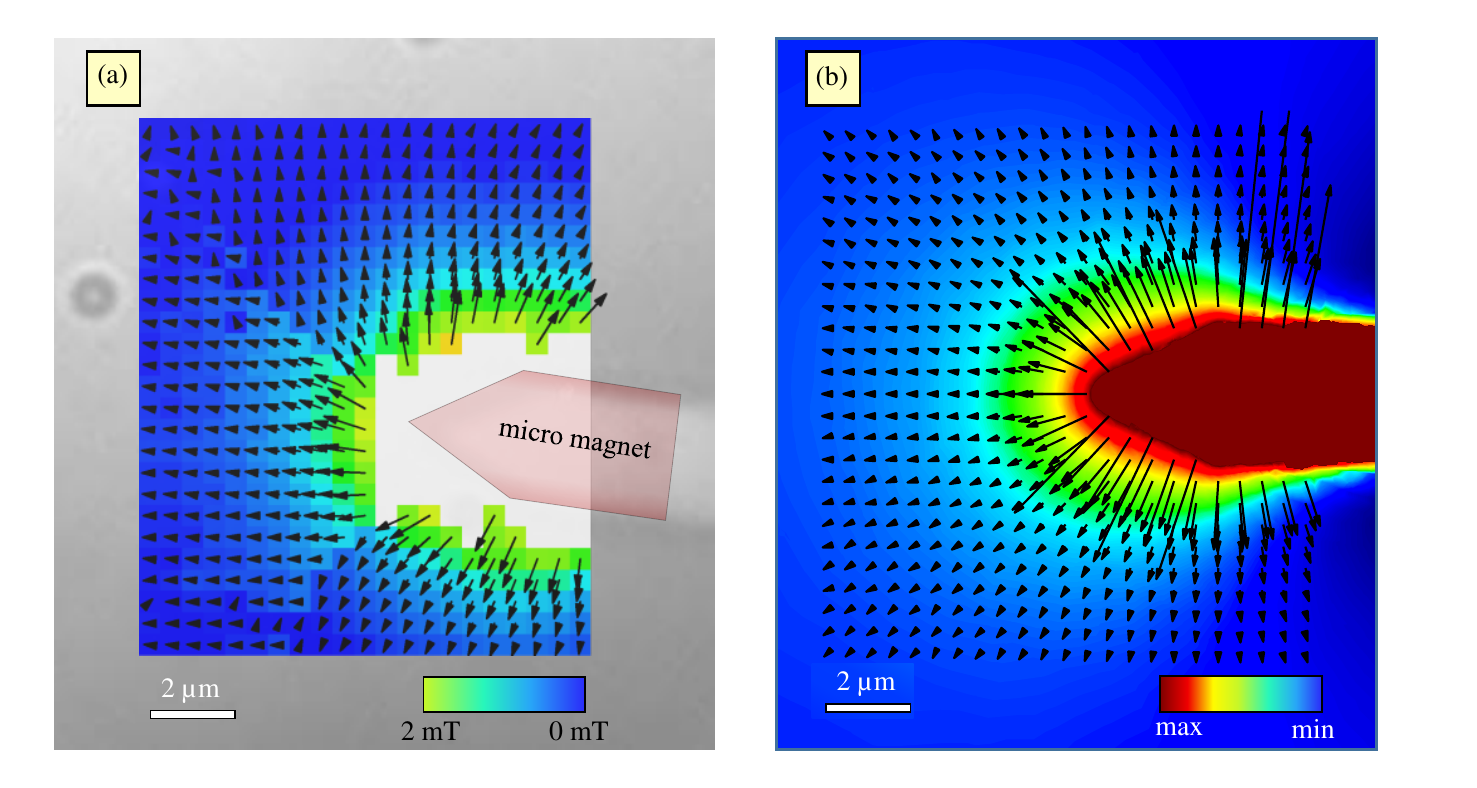}
\caption{Comparison between measured (a) and simulated (b) stray field distribution of the micromagnet. We recognize the white space in the measured data to be caused by the strong fields and field gradients at these grid points, rendering a measurement of local magnetic field unfeasible in the given configuration. The scan directions are not parallel to the symmetry axis of the structure in order to place the micromagnet at the appropriate position relative to the waveguide. Therefore the grid of the experimental results is slightly rotated. The background in part (a) is a b/w image from the diamond surface, wherein the microstructure is marked.}
\label{fig:05}
\end{figure*}

We start by identifying the angles of laser polarization that show the highest degree of selective excitation. As shown in Fig.~\ref{fig:04}~(a), these are found at $\SI{0}{\degree}$, $\SI{70}{\degree}$, and $\SI{140}{\degree}$. The deviation from the expected $\SI{60}{\degree}$ intervals arise from artifacts induced by optical elements in the setup. The beam splitter used for guiding the excitation onto the optical axis of microscope objective and detection path, which is mandatory in confocal optical setups, has slightly different reflectivities for the $s$- and $p$-components of the incident laser beam. Therefore, the beam splitter acts for certain angles of polarization as an additional rotator.

The application of external reference fields enables the determination of the orientation of the in-plane NV axes relative to the laboratory frame and the microstructure. Fig.~\ref{fig:04}~(b) indicates their relative alignment and further marks the trajectory along which we perform vector magnetometry that is shown in Fig.~\ref{fig:04}~(c).

We sample a trajectory of length of $l=\SI{12}{\micro\meter}$ in steps of $\Delta x = \SI{500}{\nano\meter}$ for each angle of the laser polarization. The low implantation depth ($d=\SI{10}{\nano\meter}$) of the NV centers assures that the emitted signal stems only from the region directly beneath the microstructure. For this reason we assume a perfect in-plane magnetic stray field and neglect the out-of-plane component in the further analysis.

Using the model function described in the previous section we determine the absolute values of $\Delta \nu$ for identified laser polarization angles. Being insensitive to sign of $\Delta \nu$,  the resulting value for the local field component along a specific NV orientation also only corresponds to the absolute value of component and remains ambiguous. Only upon the comparison of the three in-plane NV orientation, the number of possible reconstructions is reduced down to two. These vector reconstructions are identical except for their sign.

This allows us to map the stray field from our microstructure. The reconstructed field vectors are shown at the right edge of Fig.~\ref{fig:04} (c). We assume an evolving rotation of the field vector in order to resolve the sign-ambiguity. This last remaining question could be addressed by the application of circularly polarized MW fields, which selectively excite transitions between the $\ket{0} \rightarrow \ket{+1}$- and the $\ket{0} \rightarrow \ket{-1}$-states~\cite{PhysRevB.76.165205}.

We can finally expand the method to a two dimensional scan to investigate the complete stray field distribution. The result is shown in Fig.~\ref{fig:05} (a). The vector field rotates around the tip of the structure. The comparison with a numerically simulated stray field distribution of such a structure~\footnote{Simulation was performed using COMSOL Multiphysics, AC/DC-Module.} as is shown in Fig.~\ref{fig:05} (b) strongly supports the reliability of our method.

It further reveals some drawbacks of the procedure. When the observation spot is chosen close to the magnet, two problems occur. First, the strength of the local magnetic field shifts the energetic position of the $\ket{0} \rightarrow \ket{\pm 1}$-transitions out of the sampled frequency interval. Second, the field gradient along the observed defect ensemble leads to a strong broadening of the resonances in the ODMR signal. The spectra in Fig.~\ref{fig:04} (c) at the strongest energetic shift already mildly indicate these effects.

Both issues can actually be avoided. Adjustment of the sampled frequencies combined with nano-patterning of diamonds~\cite{njphausmann,111litho} will allow for further improvements of the presented principle of magnetometry. At the same time, the utilization of nano-structured diamond samples will help to increase the overall spatial resolution. 

To summarize, we demonstrated how the ambiguities of an ODMR spectrum of an ensemble of NV centers can be lifted by the application of properly polarized optical excitation. The linear polarization of the laser can reduce the number of excited NV orientation to two in case of incidence along the [111]-crystallographic axis. With help of linearly polarized MW fields, a single NV orientation can be chosen to contribute to the ODMR spectrum. The sensitivity of the method is limited by the transverse crystal field splitting E, but magnetic fields small as $\SI{50}{\micro\tesla}$ can be reliably detected. This allows for the mapping of magnetic fields as is exemplified on the stray field of a ferromagnetic microstructure. The method can be applied as well to AC magnetometry schemes where higher sensitivity limits can be reached. The temperature range in which the principle is employable extends from liquid helium temperatures~\cite{apld(t)} to at least $\SI{550}{\kelvin}$~\cite{PhysRevX.2.031001}.
 
The authors gratefully acknowledge financial support from the EU ERC-AG (Project 3-TOP). The authors further thank T. Borzenko for assistance in the lab. 

\bibliography{dissrefs}
\end{document}